\documentclass[
    final            
  ]
  {aipproc}

\layoutstyle{8x11single}

\begin{document}

\title{No indication of $f_0(1370)$ in $\pi\pi$ phase shift analyses}


\author{Wolfgang Ochs}{
  address={Max-Planck-Institut f\"ur Physik, D-80805 Munich, F\"ohringer
Ring 6, Germany}
}

\begin{abstract}

The scalar meson $f_0(1370)$ - indicated in particular in the low
energy $p\bar p \to$ 3 body reactions - is a crucial element in certain
schemes of the scalar meson spectroscopy including glueballs.  The most
definitive results can be obtained from elastic and inelastic $\pi\pi$ phase
shift analyses using the constraints from unitarity where the discrete
ambiguities can be identified and resolved.  We reconsider the phase shift
analyses for $\pi^+ \pi^- \to \pi^+ \pi^-,\ \pi^0 \pi^0,\ K \bar K,\ \eta \eta$.  While a clear
resonance signal for $f_0(1500)$ in the resp.  Argand diagrams is seen in all
channels above a large ``background'' from $f_0(600)$ there is no
clear signal of a second resonance ``$f_0(1370)$'' in this mass range
in any reaction, at the level of $\sim$10\% branching ratio into $\pi \pi$.
\end{abstract}

\maketitle


\section{ROLE OF $f_0(1370)$ IN SCALAR MESON SPECTROSCOPY} 

There is a particular
interest in scalar meson spectroscopy since the lightest gluonic meson 
is expected within QCD with these quantum numbers. The QCD predictions for the
glueball mass and a possible mixing with $q\bar q$ states is still
controversial but the mass falls into a range of 1000 - 1700 MeV. 
Therefore it is important
to identify the low mass states with their spin parity, say below 2 GeV, and to
classify them into flavour multiplets and glueballs according to their
production and decay properties. Different scenarios
have been proposed in the past for these lowest multiplets.

According to a popular scheme the states $ f_0(600)/\sigma,\
K^*(900)/\kappa,\ f_0(980),\ a_0(980)$
are grouped into the lightest nonet
either built from $q\bar q$ or $qq \bar q\bar q$.  The next higher multiplet
of $q\bar q$ type would include $a_0(1450)$ and $K^*(1430)$. For
completion of the nonet two isoscalar states are needed but 
three are found in this mass range according to 
the Particle Data Group (PDG) \cite{pdg}: $f_0(1370),\ f_0(1500)$ and $
f_0(1710)$.  This offers the possibility to introduce the scalar glueball
which together with the two isoscalar $q\bar q$ states would mix into the
observed three scalar $f_0'$s.  There are various realisations of such a
mixing scheme first considered in \cite{ac}.

A critical input in this scheme is the very existence of the three
isoscalars, so that there is one supernumerous state for a usual $q\bar q$
nonet.  However, the state $f_0(1370)$ cannot be considered as firmly
established.  This state appears with different parameters in the fits and
the PDG quotes mass and width in the ranges $M=1200-1500$ MeV and
$\Gamma=300-500$ MeV.  There are various modes quoted as ``seen'' in the
listing, but not a single number is considered to be established among
the Branching Ratios or ratios thereof despite many analyses over the years. 
Quoted results from different experiments are often in conflict.  Moreover,
we do not see a convincing direct signal, neither a peak above a low
background nor a clear resonance effect in an energy independent (point by
point) phase shift analysis.
The experimental status of $f_0(1370)$ is very different from the one of the 
nearby $f_0(1500)$ where the analyses have
converged and now there are well defined mass and width parameters with
errors <10 MeV and five well established branching ratios.  Also there are
clear direct signals from resonance bands in the Dalitz plots of 3 body decays and
also from phase shift analyses -- what is lacking for $f_0(1370)$.

There may be another problem with the above classification scheme 
in that the mass of the glueball of a full QCD calculation is lower than the
often quoted mass around 1600 MeV which refers to QCD lattice calculations
in quenched approximation. A recent full (unquenched) calculation 
in fact suggests a lower number, around 1000 MeV \cite{unquenched}, 
which is below the above 3 mass states, but more work is needed for
definitive conclusions.
Likewise, in the QCD sum rule
approach the lightest scalar glueball is found around a mass of 1000 MeV
\cite{sumrules}.

In view of the problems with  $f_0(1370)$ a classification scheme without
this state has been proposed some time ago \cite{mo}. It has been argued
that the activity in this
mass region be related to the high mass part of the broad state called now
$f_0(600)/\sigma$. Then, the states $f_0(980),\ a_0(980),\
a_0(1450),\ K^*(1430)$ are considered to form the lightest $q\bar q$
nonet while the broad low mass state is considered to be of a gluonic nature.
The ``Breit
Wigner mass'' of the broad state where the phase shift passes
through 90$^\circ$ is found around 1000 MeV, to remember we also write
$f_0(600-1000)$; the width is of size similar to the mass.  

The existence of $f_0(1370)$ therefore plays a crucial role for the
interpretation of the low mass scalar spectrum with or without glueball.
We look for an answer from a detailed phase shift analysis of elastic and
inelastic $\pi\pi$ scattering.

\section{PHASE SHIFT ANALYSIS OF $\pi\pi$ SCATTERING}

A resonance in a scattering process corresponds to a pole of the scattering
amplitude as function of the complex energy $E$ or $s=E^2$.  Such a pole
$1/(s_0-s)$ can be visualized as a closed circle in the complex amplitude 
plane for physical real $s$; in
case of a moving background a circular structure remains although with some
distortion.  Examples for the movement of the isoscalar elastic $\pi\pi$
scattering amplitude in the complex plane (``Argand diagram'') obtained from
global fits to various $\pi\pi$ data with circles from the various
resonances including $f_0(1370)$ can be seen, for example, in Refs. 
\cite{bugg}.  

In the presence of sizable background the existence of a resonance can be
demonstrated if the complex partial wave amplitude $T_\ell$ is reconstructed
from data in a sequence of mass bins and gives evidence for a 
resonance circle.  With
sufficient statistics this procedure can establish a resonance even without
explicitly fitting parametric amplitudes. 

\subsection{Data on $\pi\pi\to\pi\pi$ scattering}
\begin{figure}[h]
\begin{tabular}{@{}lll}
\includegraphics*[angle=-90,width=5cm,bbllx=3cm,bblly=1.5cm,bburx=19.5cm,
bbury=19.5cm]{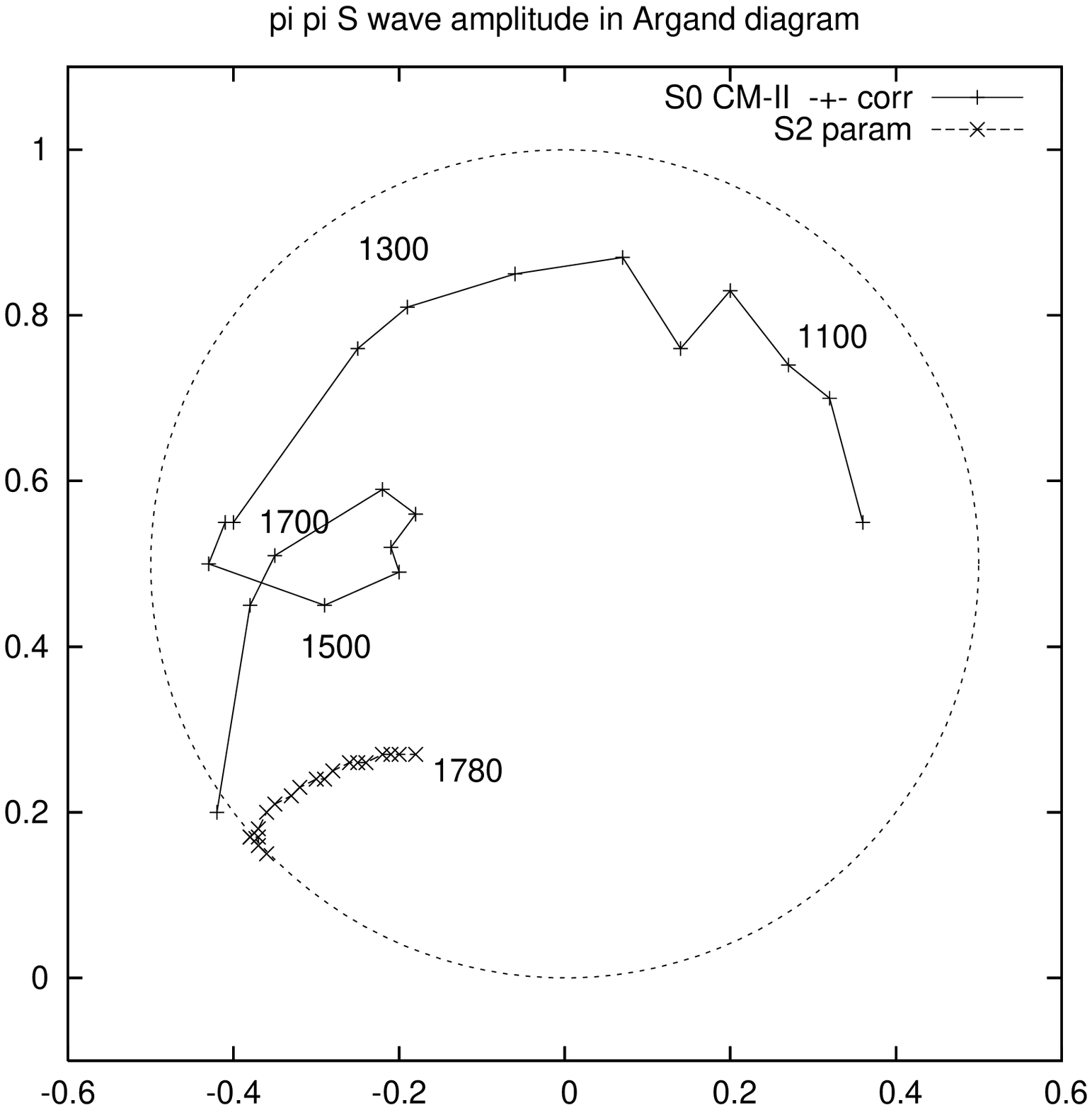}
\includegraphics*[angle=-90,width=5cm,bbllx=3cm,bblly=1.5cm,bburx=19.5cm,%
bbury=19.5cm]{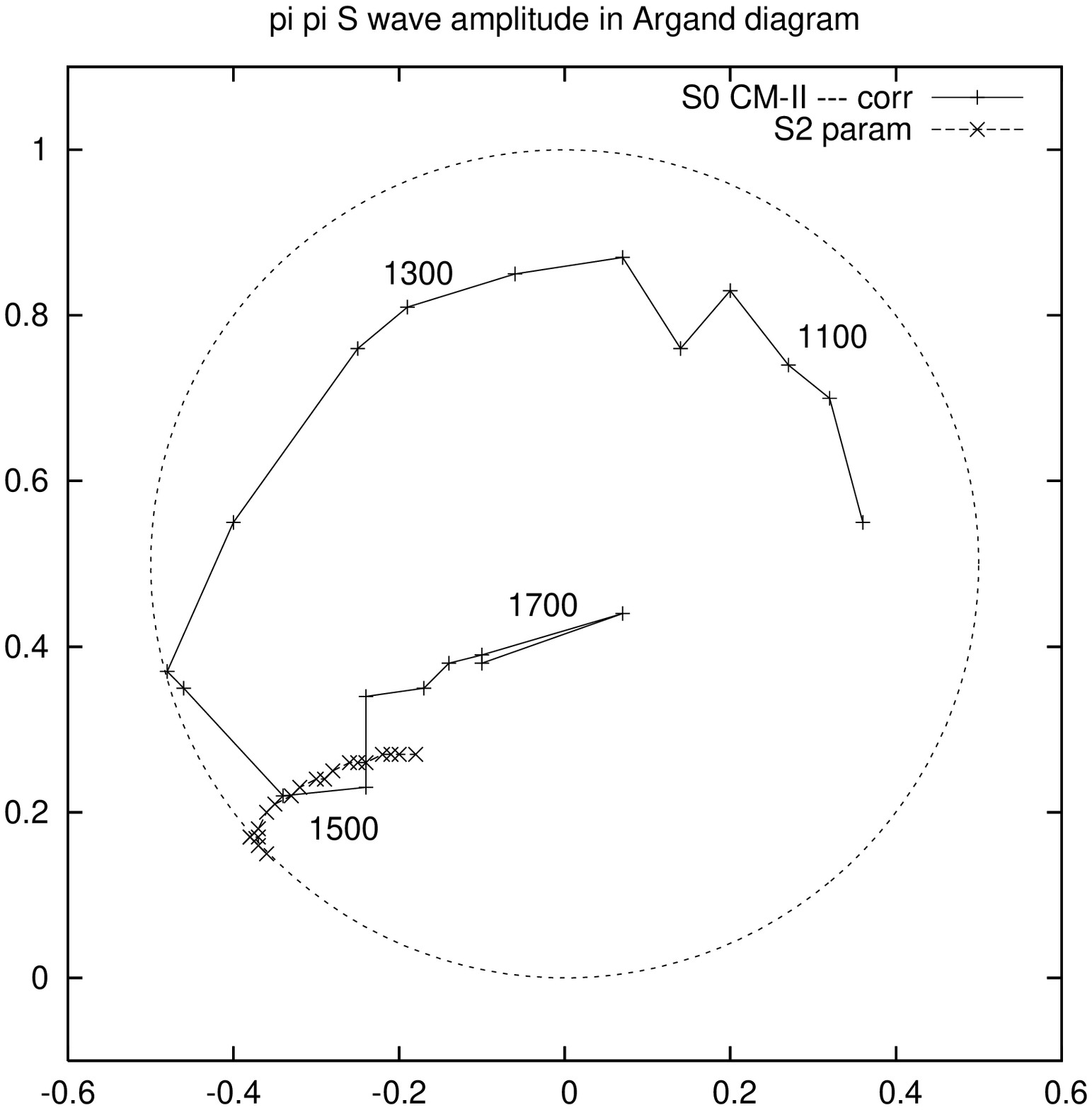}
\includegraphics*[angle=-90,width=5cm,bbllx=3cm,bblly=1.5cm,bburx=19.5cm,%
bbury=19.5cm]{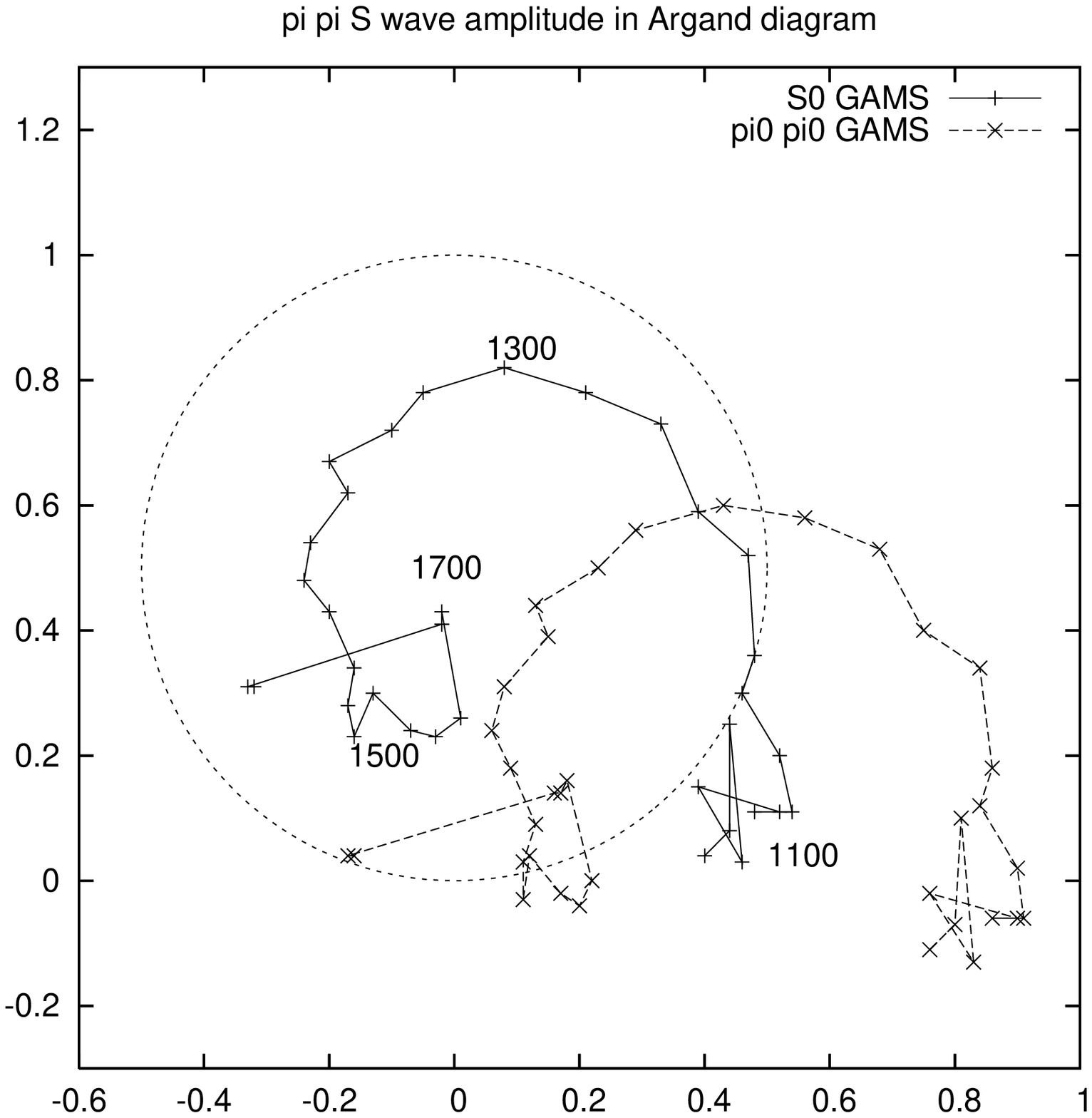} \\
\end{tabular}\\
\label{fig:arg}
\caption{Argand diagrams (Im $T_I^\ell$ vs. Re $T_I^\ell$) for isospin waves
$S_I$:
The two solutions for $S_0$ ($-+-$) and  $S_0$ ($---$) for 
$\pi^+\pi^-\to \pi^+\pi^-$ (CM-II \cite{cm2}) and for $\pi^+\pi^-\to
\pi^0\pi^0$ (GAMS \cite{gams}) and its component $S_0$ (from left to right). 
Also shown on the left panel is the isotensor component $S_2$. 
Numbers for $\pi\pi$ masses in MeV.}
\end{figure}

Such data have been obtained from the analysis of reactions like $\pi p \to
\pi\pi n$ by isolating the one-pion-exchange contribution. The measurement
of the angular distributions and their moments allows for a decomposition of
the $\pi\pi$ amplitude into partial waves up to some descrete ambiguities classified
by the ``Barrelet-zeros''. The unknown overall phase can be fixed by 
assuming a Breit-Wigner form of the leading resonances $\rho(770),\ f_2(1270)$
and $\rho_3(1690)$.
Phase shift analyses for  $\pi^+\pi^-\to\pi^+\pi^-$ became available from the 
CERN Munich experiment \cite{grayer} through different analyses:
\noindent
1. CERN Munich-I \cite{cm} with a global K matrix fit and subsequent
bin-by-bin analysis with one favoured solution;\ 
2.~CERN Munich-II \cite{cm2} with a fully energy independent analysis
resulting in 4 different solutions, two found acceptable. This 
analysis included all correlations between the angular moments 
and is carried out in larger mass
bins resulting in smaller errors for the phase shifts in comparison to
CM-I;\
3.~Similar results with 4 solutions in an analysis of CM-I data
have been obtained by Estabrooks and Martin \cite{em}. 
The two acceptable solutions by CM-II are shown in the ``Argand-diagrams''
of Fig.
\ref{fig:arg}. Also shown is the $I=2$ component used in the present analysis
which becomes inelastic and moves inside the circle above 1300 MeV
(see also Ref. \cite{ochsmont}).

An extended experiment with polarised target by the CERN-Krakow-Munich
experiment \cite{ckm} has allowed decisive tests of the underlying exchange models.
A phase shift analysis has found a unique solution for $\pi\pi$ mass 
$M<1700$ MeV but
with large errors for the parameter $\eta^0_0$ for $M>1400$ MeV.
Remarkably, the results for the phases $\delta_0^0$ from all the above analyses
agree for $M<1400$ MeV. For our present study we take the 
phase shift results from CM-II which have the smallest errors. 

There are also results on the reaction $\pi^+\pi^-\to \pi^0\pi^0$ obtained
from the corresponding $\pi p$ reaction \cite{gams,bnl}.  The GAMS data
\cite{gams} have sufficiently small errors above 1400 MeV to be compared
with $\pi^+\pi^-$ data.  
The reconstructed $\pi^0\pi^0$ amplitude as well as the $I=0$ component
are shown in the rightmost panel of Fig.  \ref{fig:arg} where the $I=0$
amplitude is found
largely inside the unitarity circle as it should be \cite{ochsmont}.
The results on the $I=0$ $\pi\pi$ amplitude obtained from charged and
neutral pion reactions should agree. One can see that 
this is true indeed for the solution ($-+-$) (leftmost in Fig. \ref{fig:arg})
at least at the qualitative level: both solutions describe a circular motion
of the amplitude with peak near 1300 MeV and with an additional 
small circle evolving towards the center
near 1500
MeV. The two amplitudes are not exactly in agreement but slightly displaced
which we relate to systematic uncertainties coming from the $I=2$
amplitude badly known in the inelastic region and from the precise 
form of the line shape of the leading resonances used to determine the
unknown overall phase.

The second, small circle is not seen in Sol. ($---$). It represents the state
$f_0(1500)$ with a diameter of about 0.3 which corresponds to the elastic width
and is in good agreement with the PDG result $(34.9\pm2.3)$\%. This
agreement is remarkable as our number comes from the imaginary part of the 
elastic $\pi\pi$ amplitude whereas the PDG value is derived from the
measurement of all channel
cross sections. 

On the other hand, there is no sign of any additional resonance circle in
the 1300 MeV range neither in the CERN Munich nor in GAMS data which would
reflect the existence of $f_0(1370)$ in the way suggested from the global
fits in \cite{bugg}. 
Assuming we should have seen a circle of size 1/3 of the one from
$f_0(1500)$ we can give an upper limit $B(f_0(1370)\to \pi\pi) < 10\%$.

\subsection{Data on $\pi\pi\to K\bar K$ scattering}

Again such inelastic processes can be obtained from the corresponding $\pi
p$ reactions. The experiments with highest statistics are:
\noindent
1. Argonne experiment on $\pi^- p\to K^-K^+ n$ \cite{argonne}; with primary
energy of 6 GeV and $\pi\pi$ mass range $M_{K\bar K}< 1.55$
GeV. Besides the amplitudes with $\pi,\ A_1,\ A_2$ exchanges contributing to 
$\pi\pi$ production
there are now also $\omega,\ B,\ Z$ exchanges and the $S$ wave can be
produced with $I=1$. Then the complete analysis yields 8 different phase
shift solutions; they are all eliminated but one.
\noindent
2. Brookhaven experiment on $\pi^- p\to K^0K^0 n$ \cite{bnlkk}; this
experiment with 3-4 times the statistics of other $K^0K^0$ experiments and
with a 23 GeV pion beam has the extended mass range  $M_{K\bar K}< 2.4$ GeV.
There are even partial waves only and the analysis yields just 
two solutions below
1.7 GeV  related by a sign ambiguity. For these reasons we compare here
with the results from this experiment. Both experiments \cite{argonne,bnlkk} 
agree roughly on the
$S$ wave magnitude but indicate a difference 
in phase at low mass $M_{K\bar K}< 1.2$
GeV.

\section{$S$-matrix fit to isoscalar $S$-wave}

Next we fit the discussed data from CM-II and BNL with a
parametrisation of the coupled channel $S$ matrix in the range
$1.0 <M_{\pi\pi}<1.7$ GeV together with data from CM-I from the range
$0.7 <M_{\pi\pi}<1.0$ GeV including the resonances
\begin{equation} 
 \sigma/f_0(600-1000),\ f_0(980),\ f_0(1500),\ [f_0(1710)]
\label{res3}
\end{equation}
where we also take into account the tail of $f_0(1710)$ with parameters from PDG but
free coupling strength.

A popular ansatz is based on the addition of several $K$ matrix poles, but
the relation to the $T$ matrix poles is not obvious and one cannot easily
switch on or off a particular resonance.  Also it is not obvious why K
matrix poles should be added, in the elastic case $K=tg \delta$.  We prefer
using a model where the phases are added in the elastic region which is
obtained if for two resonances the corresponding $S$ matrices are multiplied
$S=S_1 S_2$ which may be generalised to one inelastic and one elastic
resonance (or background) according to \cite{michael}.  This leads for the
combined $T$ matrix to the well known expression $T=T_2+T_1
\exp(2i\delta_2)$.  We generalise this to an Ansatz for inelastic
resonances: the broad ``background'' from $f_0(600-1000)$ interfering with
narrow resonances $f_0(980)$ and $f_0(1500)$ MeV
\begin{equation}
S=S_1 S_2 S_3 \ldots
\label{sdef}
\end{equation} 
We assume here simple Breit-Wigner forms for the resonances $i$ coupling to
channels $k,\ell$
\begin{equation}
S_i=1+2iT_i\quad  {\rm where} 
\quad T_{i,k\ell}\ =
     \displaystyle\frac{M_0 \Gamma_k(M)\Gamma_\ell(M)} 
{M_0^2-M^2-i M_0\Gamma(M)}
\label{tdef}
\end{equation}

We perform a common fit of this parametrization to the experimental data in 
$\pi\pi$ and $K\bar K$ channels for the parameters
inelasticity $\eta$ and phase $\delta$ for the amplitude of $\pi\pi\to
\pi\pi$ by $T_{11}=(\eta \exp(2i\delta)-1)/2i$ 
and to $|S|^2$ and $\phi$ 
for $\pi\pi\to K\bar K$ by $T_{12}=|S|\exp(i\phi)$.
We include in the fit the 3 (+1) resonances in (\ref{res3})
each one with parameters $M$, $\Gamma$ and decay fractions
$x_i=\Gamma_i/\Gamma,\ \Gamma=\sum \Gamma_i$
where appropriate line shapes for the channels $\pi\pi,\ K\bar K$ and
$4\pi$ are included.  The $4\pi$ channel is kept to satisfy unitarity
but without experimental input.

\begin{figure}[t]
\begin{tabular}{ccc}
\includegraphics*[angle=-90,width=5cm,bbllx=3cm,bblly=6cm,bburx=19.0cm,
bbury=23.0cm]{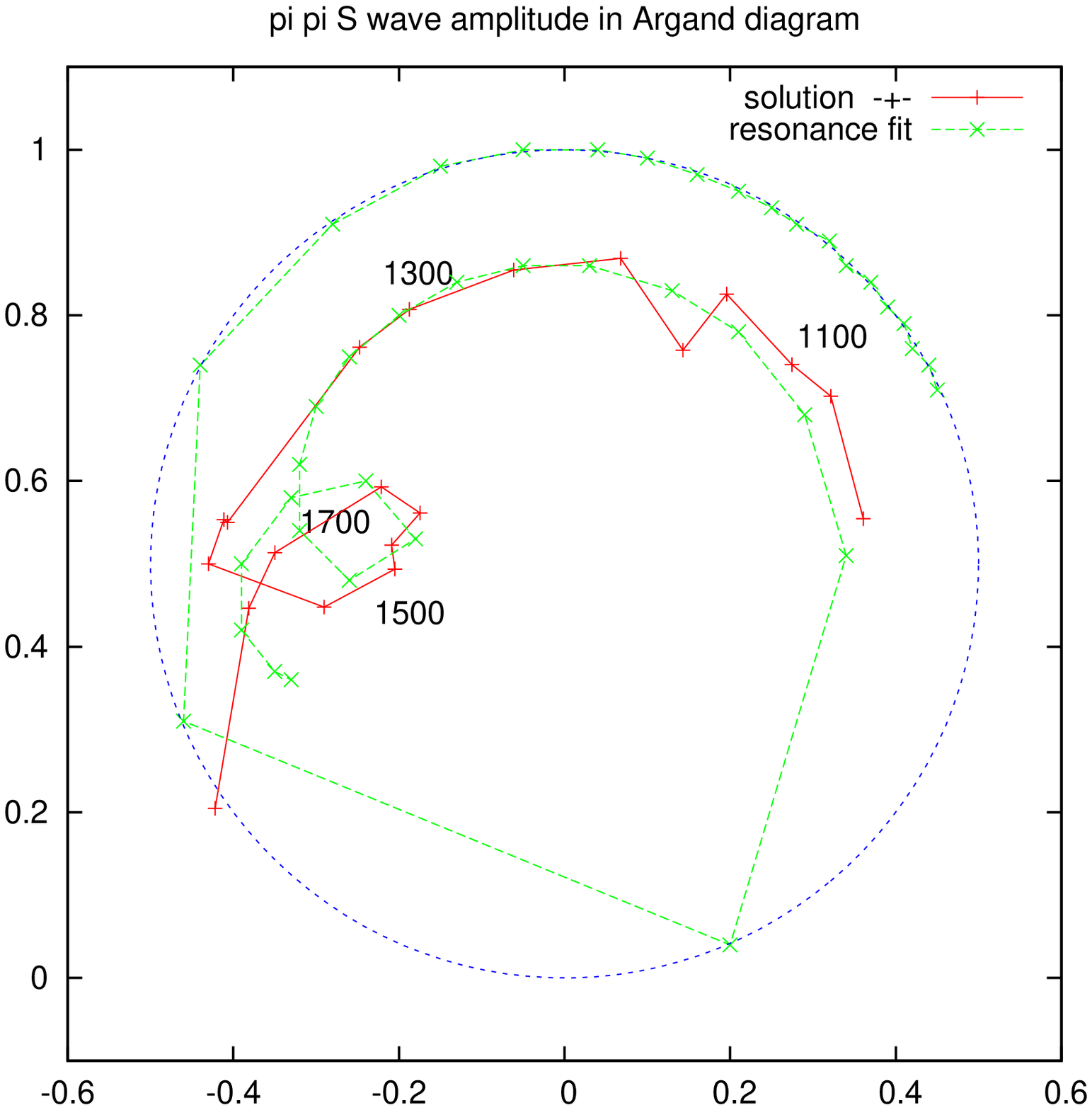}&
\includegraphics*[width=5cm,angle=-90,bbllx=3cm,bblly=6cm,bburx=19.0cm,%
bbury=23.cm]{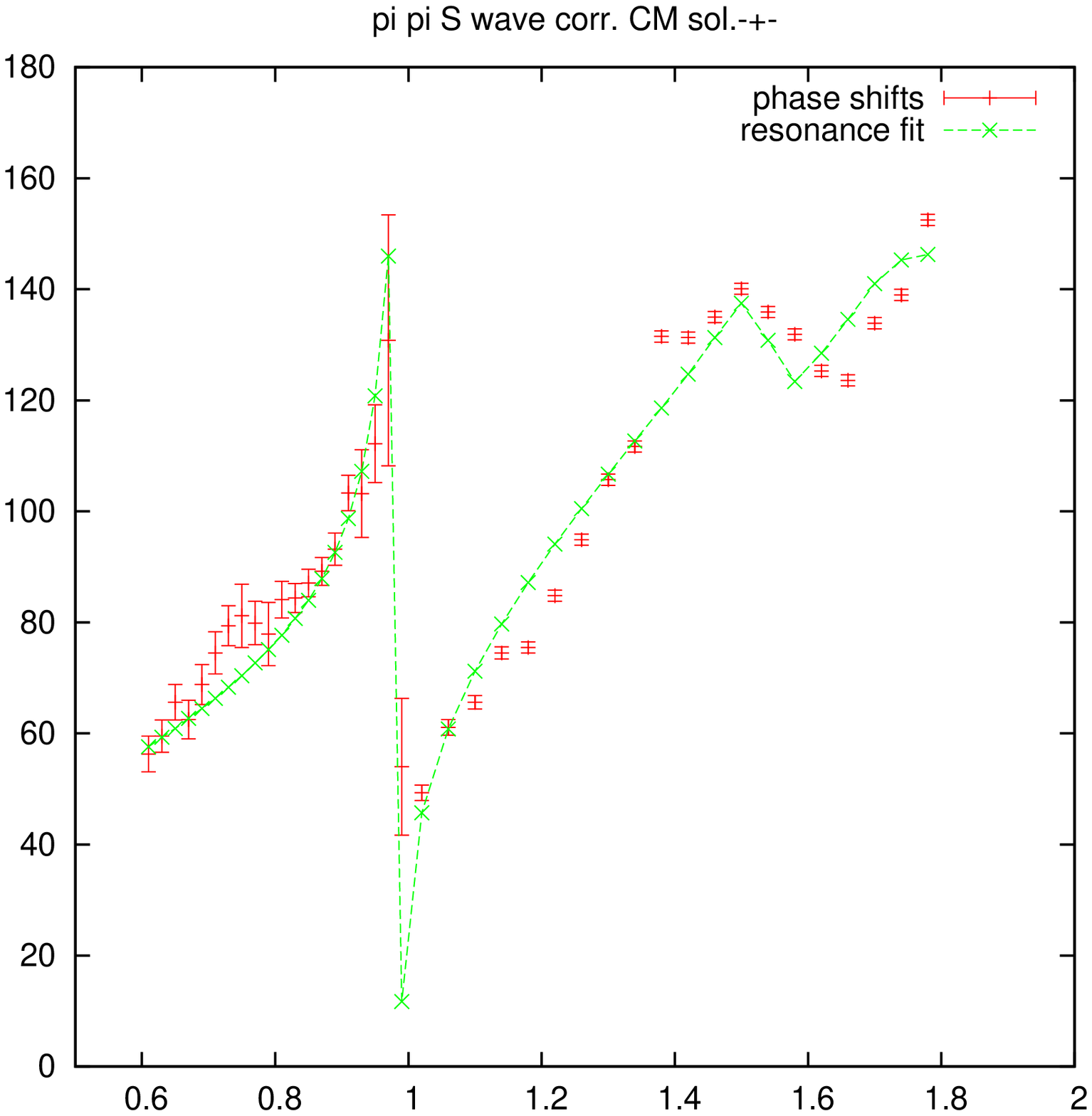} &  
\includegraphics*[width=5cm,angle=-90,bbllx=3cm,bblly=6cm,bburx=19.0cm,%
bbury=23cm]{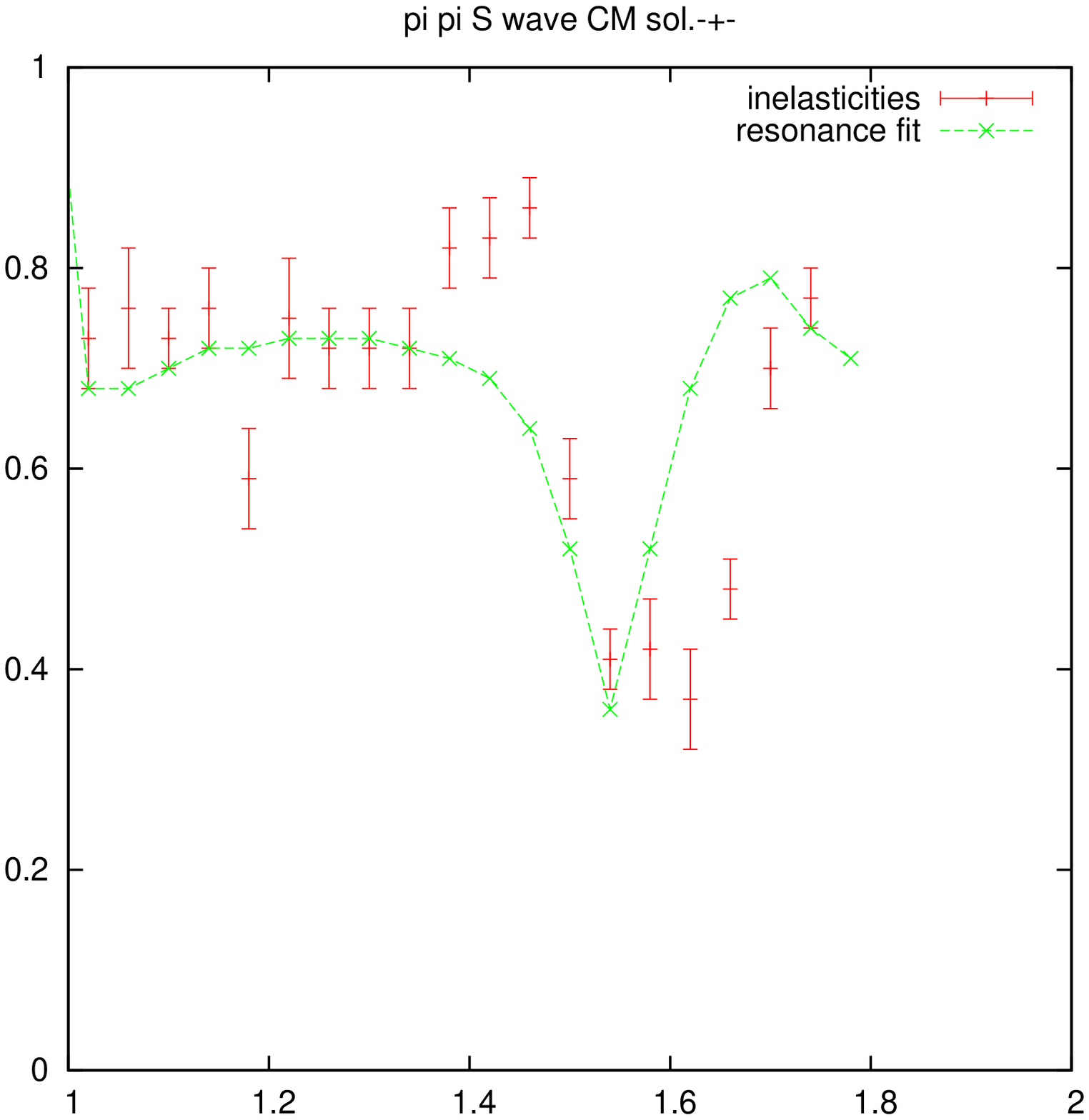}  \\
\end{tabular}
\label{fig:fit}
\caption{Common fit to $\pi\pi$ and $K\bar K$ data: results for 
Argand plot (Im $S_0$ vs. Re $S_0$), phase shifts
$\delta_0^0$ in degrees and inelasticities $\eta_0^0$ for $S$-wave $\pi^+\pi^-\to \pi^+\pi^-$.}
\vspace{-0.3cm}
\end{figure}

The results for elastic $\pi\pi$ scattering are presented in Fig. \ref{fig:fit}.
It can be observed that the main features of the data are reproduced: the slowly
moving background whith phase shifts passing 90$^\circ$ near 1000 MeV, interrupted
by the strong variation from $f_0(980)$ and a second rapid variation near 1500 MeV.
One can see from the Argand diagram that the fitted circle near 1500 MeV 
is a bit rotated
away from the circle in the data which leads to a shift of the dip in $\eta$
away from the experimental value. We assume that this is a shortcoming of
our rather simple parametrisation with minimal resonances and no further
``background''. However, it does not appear justified to us to include 
an additional resonance to describe
the effect which would generate an additional circle. 
The amplitude (both $\eta$ and $\delta$) behaves smooth near 1300
MeV without any sign of an additional $f_0(1370)$.

\begin{figure}[h]   
\begin{tabular}{ccc}
\includegraphics*[angle=-90,width=5cm,bbllx=3cm,bblly=6cm,bburx=19.5cm,
bbury=23.5cm]{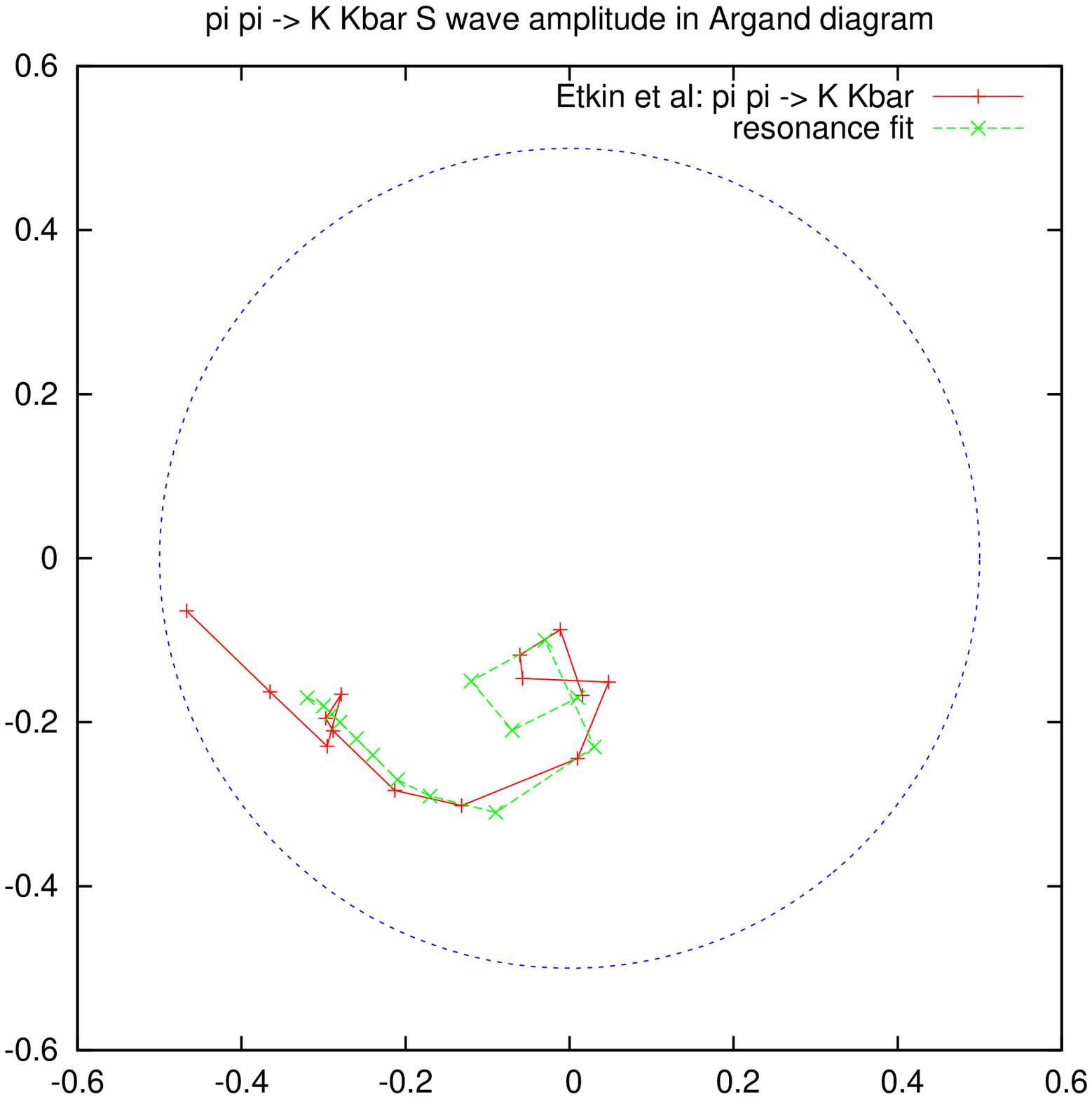}  &
\includegraphics*[width=5cm,angle=-90,bbllx=3cm,bblly=6cm,bburx=19.5cm,%
bbury=23.5cm]{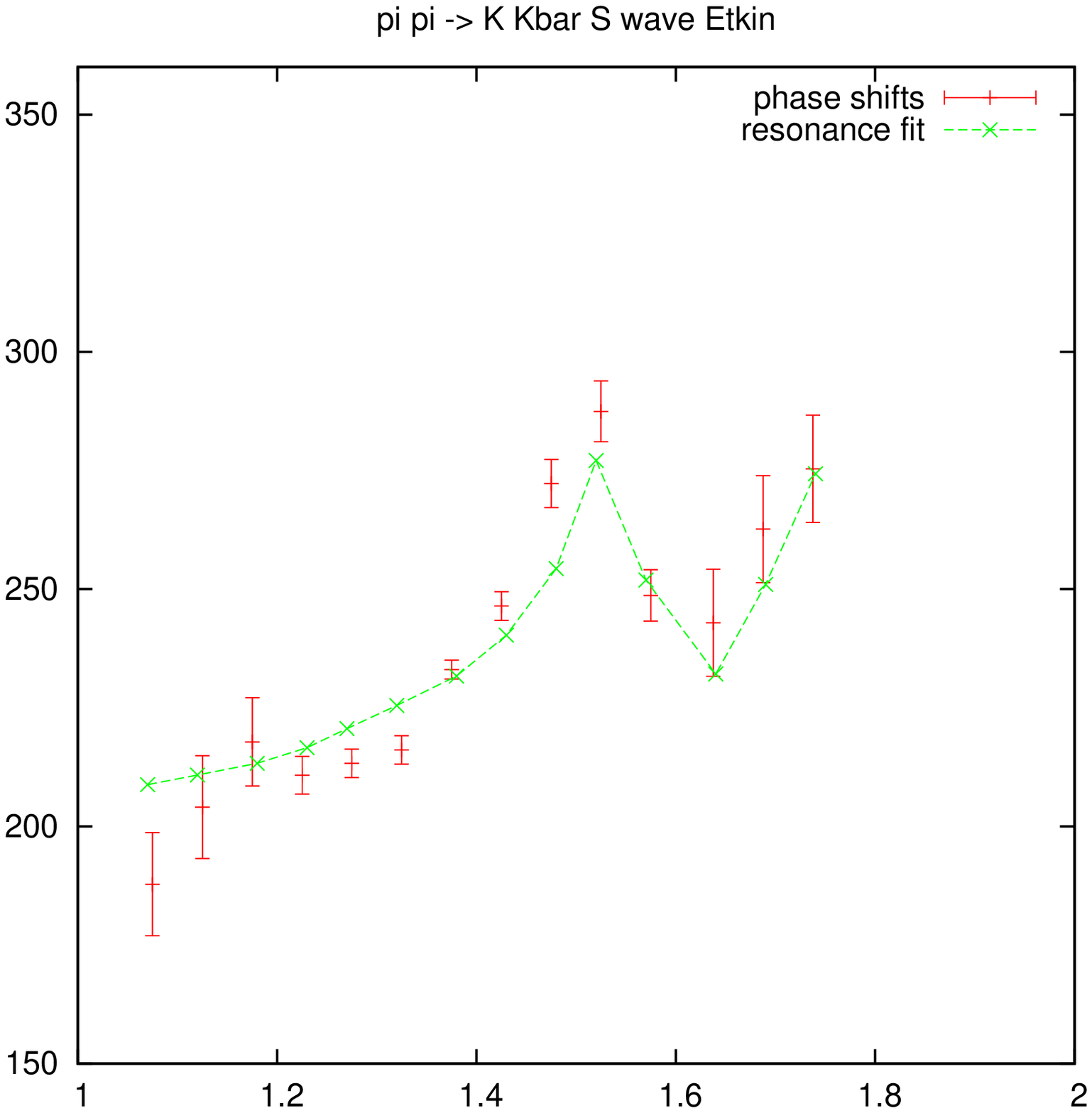} &  
\includegraphics*[width=5cm,angle=-90,bbllx=3cm,bblly=6cm,bburx=19.7cm,%
bbury=23.5cm]{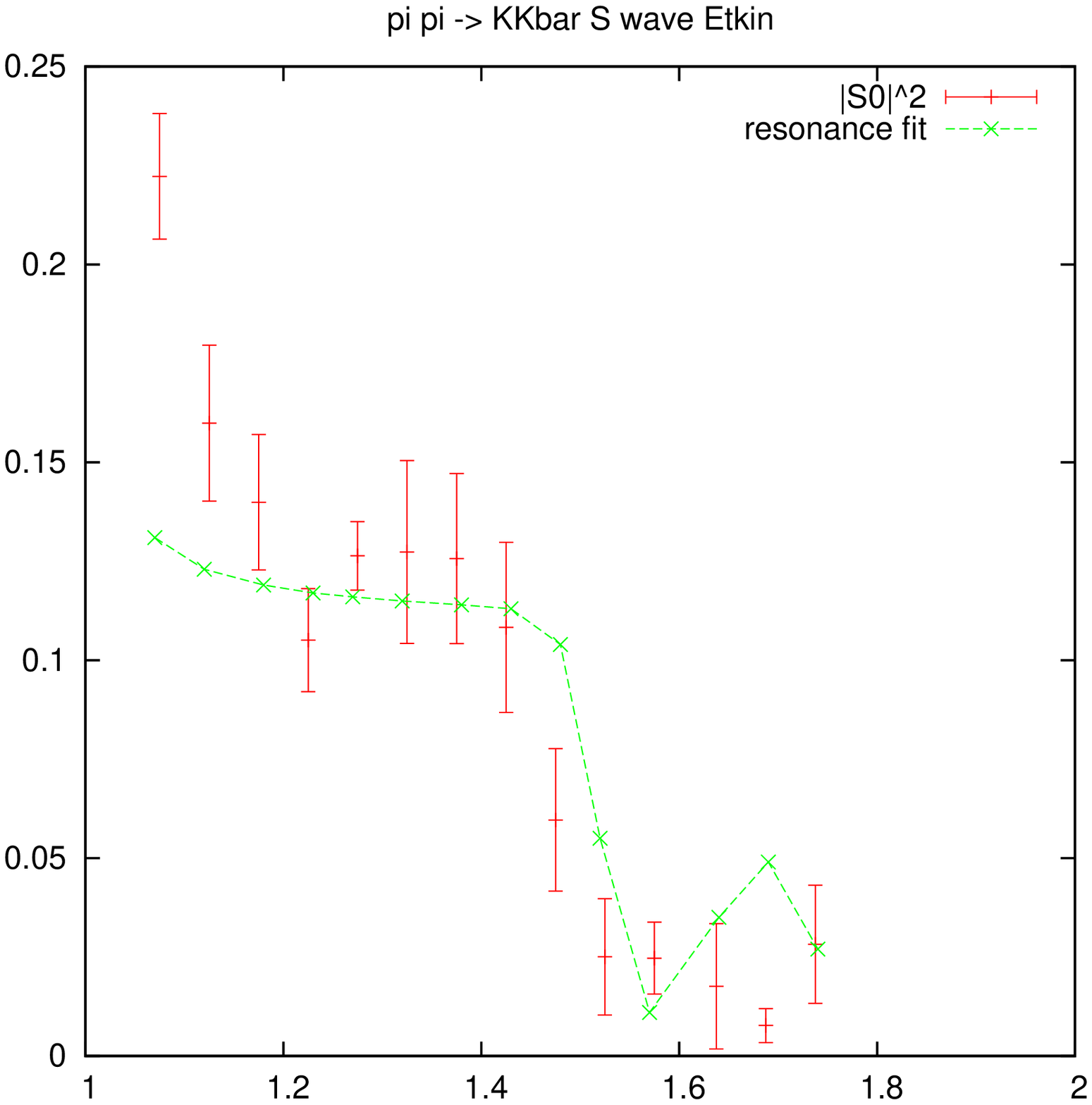} \\
\end{tabular}  
\label{fig:kk}
\caption{Common fit to $\pi\pi$ and $K\bar K$ data: results for
Argand plot, phase shift
and magnitude $|S|^2$ for $S$-wave $\pi^+\pi^-\to K\bar K$.}
\end{figure}

Moving on to the $K\bar K$ channel in Fig. \ref{fig:kk} we note again a
good overall agreement with a rapid variation near 1500 MeV in both
magnitude and phase of the amplitude and a smooth behaviour otherwise. The
discrepancy in magnitude near threshold cannot be resolved without
disturbing the agreement with magnitude ($\eta_0^0$) in the elastic channel and the fit
represents a compromise. The inelastic amplitude has been fitted including 
an arbitrary overall phase.

There are also data on the reaction $\pi^+\pi^-\to \eta \eta$ \cite{binon} 
for which the Argand diagram has been constructed \cite{mo}. Again, there is
a nice ``small'' circle near 1500 MeV above a slowly moving background but no
indication of a second circle. These data are not yet included in the fit.

The shortcomings of the above fits in certain details suggest some
improvements.  First, it should be noted that the parametrisation
(\ref{sdef}) is to be considered as a phenomenological ansatz; it has
correct limits into elastic or mixed elastic-inelastic channels but in its
general form it is unitary but not necessarily symmetric as it should be. 
While our application is similar to the mixed case a formula with all
constraints fulfilled would be desirable.  It may be interesting to
compare with a $K$ matrix ansatz as well despite its other shortcomings.  Also one
may test resonance formulae beyond the minimal form (\ref{tdef}) and the role of
possible additional background.

\section{Summary and desirable future studies}
Using data from $\pi^+\pi^-$ elastic and charge exchange scattering
and some information on the $\pi\pi$ $I=2$ channel we are able to
find a unique solution for the isoscalar $\pi\pi$ $S$ wave. The respective Argand 
diagram clearly shows the small circle from $f_0(1500)$ consistent with
known elasticity above the broad background from $f_0(600-1000)$ (its
directly observable Breit-Wigner mass is $\sim$1000 MeV).
A simple parametric fit reproduces the main
features of the data in terms of the broad $f_0(600-1000)$ and the narrow 
$f_0(980)$ and $f_0(1500)$ while there is no direct signal from $f_0(1370)$
which yields the limit $B(f_0(1370)\to \pi\pi)<10\%$.

The main difference to the earlier studies with positive evidence for 
$f_0(1370)$
is our technique of using an energy independent analysis of the scattering
data. In this way the extraction of partial wave amplitudes and their
parametric representation can be discussed separately. Note that the phase
shift data as in Fig. \ref{fig:arg},\ref{fig:fit} together with those from 
the other partial
waves come from a very good fit to the original experimental moments with
small $\chi^2<1$ in all bins. Therefore there is very little room for additional
resonance circles. A global fit to the original data (moments,
Dalitz plot) typically has a rather large $\chi^2/ND>2$ and this
may be due to a simplified parametrisation (as in our fit) or also to an
inadequate amplitude structure, the latter case is avoided in performing a
point by point phase shift analysis first.  

It will therefore be desirable to perform energy independent analyses with
other data of high statistics. Such developments have
started for 3-body final states recently \cite{meadow}, noteworthy the
recent amplitude determination in $D_s\to 3\pi$ 
by BaBaR \cite{babar}. Such
data may be precise enough to resolve in a model independent way small
circles, i.e. resonances with small couplings, such as the suspected
$f_0(1370)$. Similar studies could be
performed for $p\bar p\to$~3 body, which needed in addition an
assumption on nucleon spin couplings. While such analyses cannot determine
absolute branching ratios from a particular channel as we can do 
in our analysis, they can help to establish the
existence of such states like $f_0(1370)$ if a positive resonance 
signal is detected.


\begin{theacknowledgments}
I would like to thank Peter Minkowski for many discussions and the 
collaboration on various aspects of this work. 
\end{theacknowledgments}




\bibliographystyle{aipproc}   

\bibliography{sample}

\IfFileExists{\jobname.bbl}{}
 {\typeout{}
  \typeout{******************************************}
  \typeout{** Please run "bibtex \jobname" to optain}
  \typeout{** the bibliography and then re-run LaTeX}
  \typeout{** twice to fix the references!}
  \typeout{******************************************}
  \typeout{}
 }

\end{document}